\documentclass[11pt]{article}
\usepackage{moriond,epsfig}

\def\Journal#1#2#3#4{{#1} {\bf #2}, #3 (#4)}

\def\citeand{\hspace*{-1pt}$^,$\hspace*{3pt}}
\newcommand{\msun}{{\rm M}_{\odot}}
\newcommand{\rsun}{{\rm R}_{\odot}}
\newcommand{\ltap}{\mathrel{\hbox{\rlap{\lower.55ex \hbox {$\sim$}}
                   \kern-.3em \raise.4ex \hbox{$<$}}}}
\newcommand{\gtap}{\mathrel{\hbox{\rlap{\lower.55ex \hbox {$\sim$}}
                   \kern-.3em \raise.4ex \hbox{$>$}}}}

\begin{document}
\vspace*{4cm}
\title{GRAVITATIONAL WAVES FROM DOUBLE WHITE DWARFS}

\author{ G. NELEMANS, S.F. PORTEGIES ZWART\footnote{SPZ is a Hubble Fellow}  AND F. VERBUNT}

\address{Astronomical Institute ``Anton Pannekoek'', University of Amsterdam\\
Boston University/MIT \\ Astronomical Institute, Utrecht University}

\maketitle\abstracts{Double white dwarfs could be important sources for space
  based gravitational wave detectors like OMEGA and LISA. We use
  population synthesis to predict the current population of double
  white dwarfs in the Galaxy and the gravitational waves produced by
  this population.  We simulate a detailed power spectrum for an
  observation with an integration time of $10^6$\,s. At frequencies
  below $\sim$\,3\,mHz confusion limited noise dominates. At higher
  frequencies a few thousand double white dwarfs are resolved
  individually.  Including compact binaries containing neutron stars
  and black holes in our calculations yields a further few hundred
  resolved binaries and some tens which can be detected above the
  double white dwarf noise at low frequencies.  We find that binaries
  in which one white dwarf transfers matter to another white dwarf are
  rare, and thus unimportant for gravitational wave detectors.  We
  discuss the uncertainties and compare our results with other
  authors.}

\section{Introduction}

Since most stars evolve into a white dwarf and since most stars are members
of binaries, close binaries containing two white dwarfs are expected to
be very numerous in our Galaxy. Such binaries may dominate the
gravitational wave signal detected by space based detectors like LISA
and OMEGA\,\cite{eis87}\citeand \cite{hbw90}. The expected numbers of
these binaries could be so large that they form a confusion limited
noise in the detector. 

\begin{figure}[t]
\begin{minipage}[t]{0.35\textwidth}
\caption{Example of the formation of a double white dwarf. The masses
  of both stars and the orbital period at each evolutionary phase are
  indicated.  The initial binary consists of a 2.0\,$\msun$ star and a
  1.9\,$\msun$ star in a 232\,day orbit (a semi-major axis of 250
  $\rsun$) (a). The more massive star evolves first into a giant and
  starts transferring matter to its companion (b). Most of the
  transferred matter leaves the binary and takes angular momentum with
  it. The orbit first shrinks but expands when the donor has become
  (much) less massive than its companion. When the donor has lost all
  its envelope it becomes a 0.42\,$\msun$ white dwarf (c).
  Subsequently the companion evolves and engulfs the white dwarf. The
  core of the companion and the white dwarf move in a common envelope
  in a rapidly shrinking orbit (``spiral-in'') until the envelope is
  ejected (d - e). A close double white dwarf results (e). The last
  phase is plotted again on a 100 times larger scale.}
\end{minipage}
\begin{minipage}[t]{0.62\textwidth}
\epsfig{figure=scenario2.ps,width=\textwidth,clip=true,angle=-90}
\psfig{figure=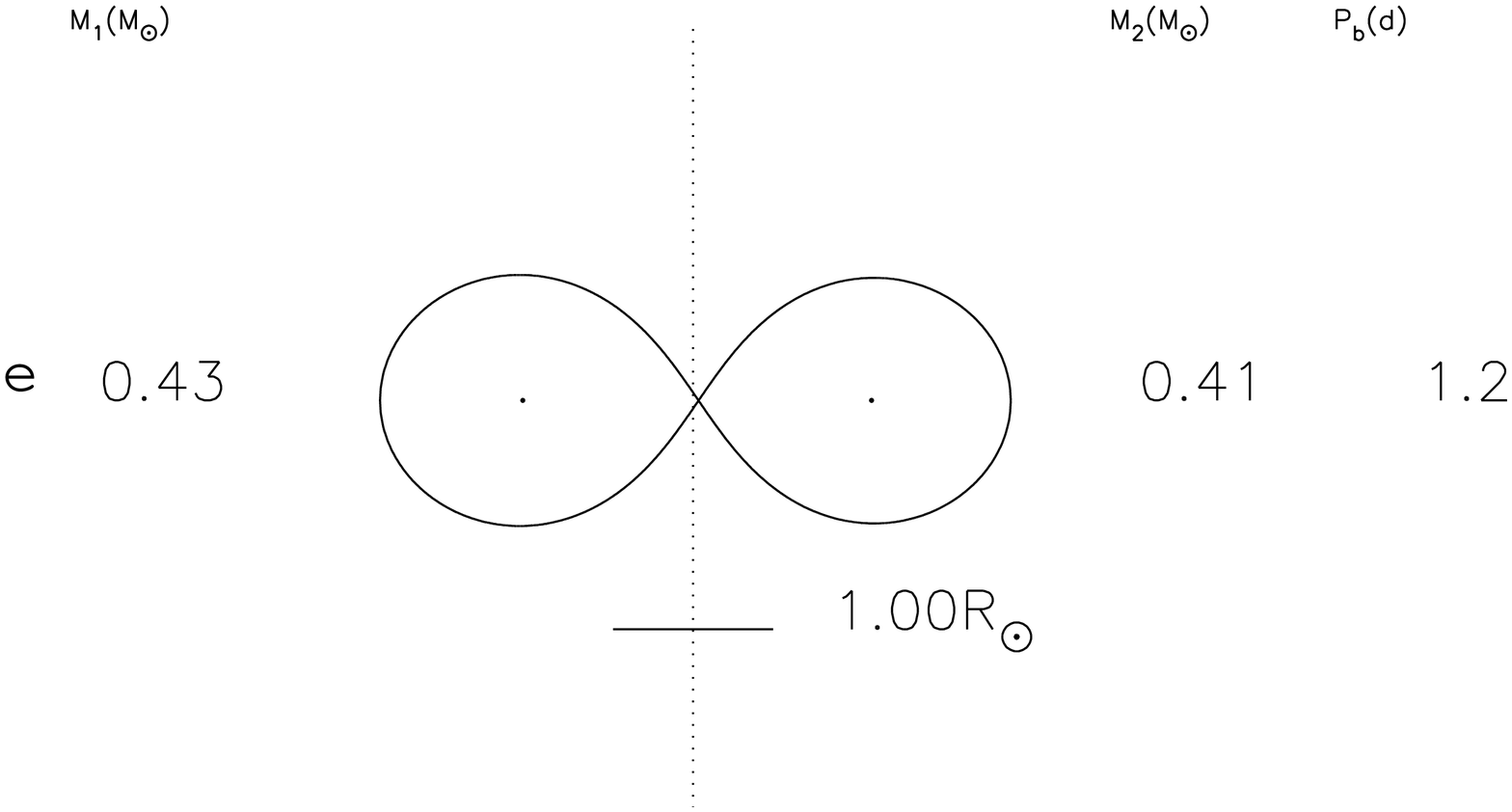,angle=-90,clip=,width=0.955\textwidth}
\end{minipage}
\label{fig:scenario}
\end{figure}

Figure~\ref{fig:scenario} shows a typical example of a binary
evolution leading to a double white dwarf. In this example the first
phase of mass transfer is stable and the second mass transfer phase
results in a common envelope and a dramatic decrease of the orbital
period.  Further evolution of double white dwarfs is driven by the
emission of gravitational waves and may lead to systems in which the
lighter white dwarf transfers matter to the more heavy white
dwarf. This is only stable when the mass ratio of the two white dwarfs
is smaller than~2/3. Double white dwarfs with stable mass transfer are
called interacting double white dwarfs or AM CVn stars.

Until recently very few double white dwarfs were known. However,
improved observations have lead to a dramatic increase in the number
of observed systems\,\cite{mar95}\citeand \cite{mdd95} (see Table
1). In section 2 we describe a detailed calculation of the number of
double white dwarfs in the Galaxy and compare their properties with
those of the observed systems. In section 3 we compute the
gravitational wave signal of these binaries and in section 4 we
compare our results with those of other authors.

\section{Population Synthesis}

To compute the current population of close double white dwarfs we use
population synthesis as follows. We initialize a large number
of `zero-age' binaries and evolve these binaries according to
simplified prescriptions for stellar and binary evolution, including
stellar wind, mass transfer (which may involve loss of mass and
angular momentum from the binary), common envelope and
supernovae. The code is improved with respect to an earlier
version\,\cite{pv96}\citeand \cite{py98} in a more accurate treatment
of the formation of white dwarfs\,\cite{npv99}.

For each initial binary the mass of the more massive component ($M$),
the mass ratio ($q \equiv m/M \le 1$, where $m$ is the mass of the
less massive component), the period ($P$) and eccentricity ($e$) are
chosen randomly from distributions given by
\begin{eqnarray}\label{eq:init}
{\rm Prob}(M) \propto & M^{-2.5}  &  {\rm for} \; 0.8 \leq M \leq 100 \, \msun \nonumber \\ 
{\rm Prob}(q) \propto & {\rm cnst} & {\rm for} \; \; \; \; 0 \leq q \leq 1  \nonumber \\
{\rm Prob}(P) \propto & P^{-1} & {\rm for} \; \; \; \; 1 \leq \log P \leq 10^6 \\ 
{\rm Prob}(e) \propto & 2 e & {\rm for} \; \; \; \; 0 \leq e \leq 1\nonumber 
\end{eqnarray}

We initialised 100,000 binaries and computed their evolution. Almost
27,500 of them became double white dwarfs with periods below 100 days.
All these binaries have circular orbits due to the strong tidal forces
during mass transfer phases. Most double white dwarfs contain two
helium white dwarfs (49\%), 25\% contains two C/O white dwarfs, 22.5\%
has both varieties and 3.5\% contains an O/Ne/Mg white dwarf. 52\% of
all systems are close enough that mass transfer starts within a Hubble
time. In 1.4\%\ of the cases the mass transfer is stable and an AM CVn
system is formed. The remaining systems merge.

The results are normalised to the estimated star formation history of
the Galaxy (4\,$\msun$\,yr$^{-1}$), assuming all stars are formed in
binaries. With this normalization the birth rate for double white
dwarfs is $0.1$\,yr$^{-1}$ in the galaxy and the merger rate is 5
10$^{-2}$\,yr$^{-1}$. Of these mergers about 10\% has a total mass of
more than the Chandrasekhar limit (1.44\,$\msun$), yielding a merger
rate of 5\,10$^{-3}$\,yr$^{-1}$. Since some models for type Ia
supernovae involve merging of these white dwarfs this can be compared
to the estimated supernova Ia rate\,\cite{vt91} of
3-4\,10$^{-3}$\,yr$^{-1}$. The birth rate of interacting white dwarfs
is only 1.3\,10$^{-3}$\,yr$^{-1}$.

\begin{figure}[t]
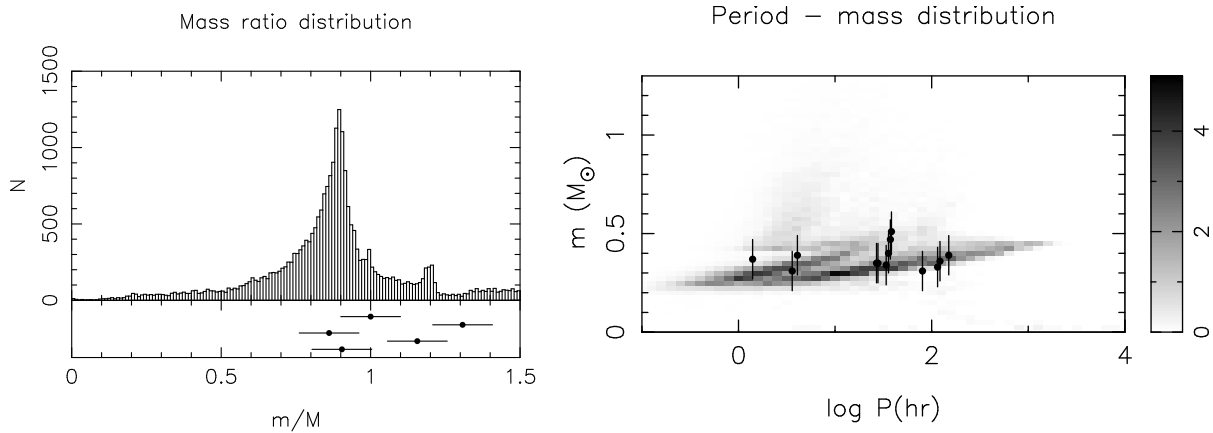

\begin{minipage}{0.39\textwidth}
\psfig{figure=q3.ps,width=1.22\textwidth,bbllx=0pt,bblly=384pt,bburx=565pt,bbury=815pt,clip=}
\end{minipage}
\hspace*{0.065\textwidth}
\begin{minipage}{0.55\textwidth}
\psfig{figure=periods3.ps,width=\textwidth,bbllx=30pt,bblly=55pt,bburx=565pt,bbury=392pt,clip=}
\end{minipage}
\caption{Left: predicted mass ratio distribution of the current double white
  dwarf population. We define the mass ratio for the double white
  dwarf as the mass of the last formed (i.e. the potentially visible)
  white dwarf over the mass of the first formed white dwarf. Right:
  gray scale plot of distribution over orbital period and mass of the
  visible component. Grey scale gives the expected number of visible
  systems with magnitude smaller than 15. Points with error bars
  indicate observed systems given in Table 1.}
\label{fig:population}
\end{figure}

We assume a constant star formation history in the galaxy for the last
10 Gyr. This gives a current population of $\sim\,3\,10^{8}$ double
white dwarfs in the galaxy at present. Figure~\ref{fig:population}
shows the mass ratio distribution and the distribution over period and
mass of the last formed (i.e. potential visible) white dwarf. The
systems currently known are indicated. We also list the properties of
these systems in Table 1.

\begin{table}[t]
\caption{Parameters of known double white dwarfs (for references see
Maxted \&\ Marsh, 1999 and Moran, 1999), with $m$ denoting the mass
of the visible (i.e. last formed) white dwarf. The mass ratio for the
double white dwarfs is defined as the mass of the last formed
(i.e. the potentially visible) white dwarf over the mass of the first
formed white dwarf.}
\begin{center}
\begin{tabular}{|p{3cm}|r|r|r||p{3cm}|r|r|r|r|} \hline
WD & $P($d) & $q$  & $m/\msun$  & WD & $P($d) & $q$ & $m/\msun$ \\ \hline
0135$-$052 & 1.556 & 0.90 & 0.47 & 0136$+$768 & 1.407 & 1.31 & 0.34 \\
0957$-$666 & 0.061 & 1.14 & 0.37 & 1022$+$050 & 1.157 &      & 0.35 \\
1101$+$364 & 0.145 & 0.87 & 0.31 & 1202$+$608 & 1.493 &      & 0.40 \\
1204$+$450 & 1.603 & 1.00 & 0.51 & 1241$-$010 & 3.347 &      & 0.31 \\
1317$+$453 & 4.872 &      & 0.33 & 1713$+$332 & 1.123 &      & 0.38 \\
1824$+$040 & 6.266 &      & 0.39 & 2032$+$188 & 5.084 &      & 0.36 \\
2331$+$290 & 0.167 &      & 0.39 &  & & & \\ \hline
\end{tabular}
\end{center}
\end{table}

We see that the predicted distributions match the (small) observed
sample well. However, since we have not modelled the selection effects
of the observed sample, we can not yet compare the {\em number} of
detected systems with the predictions.



\section{Gravitation waves}

To compute the gravitational wave signal for the population of double
white dwarfs, we distribute our ($\sim\,3\,10^{8}$) double white
dwarfs in the Galaxy according to a disk distribution with radius of
15 kpc, with the sun at 8.5 kpc from the center. The height of the
disk is 200 pc. For each of these double white dwarfs we compute
the strength of the gravitational wave amplitude from\,\cite{eis87}
\begin{equation}
h = 5.1 \; 10^{-22} \left ( \left[ \frac{M \; m}{M + m} \right] \Big/ \msun \right) \left (
\frac{M + m}{\msun} \right)^{2/3} \left ( \frac{P_{orb}}{\rm 1
\; hr} \right)^{-2/3} \left ( \frac{d}{\rm 1 \; kpc} \right)^{-1}
\end{equation}

The frequency of the wave is given by $f = 2/P_{orb}$. We add the
amplitudes together in frequency bins of $\Delta f = 1/T$ ($T =$ integration
time) to simulate the power spectrum for this population of binaries
as would be detected by gravitational wave detectors in space like
LISA and OMEGA.

\begin{figure}[t]
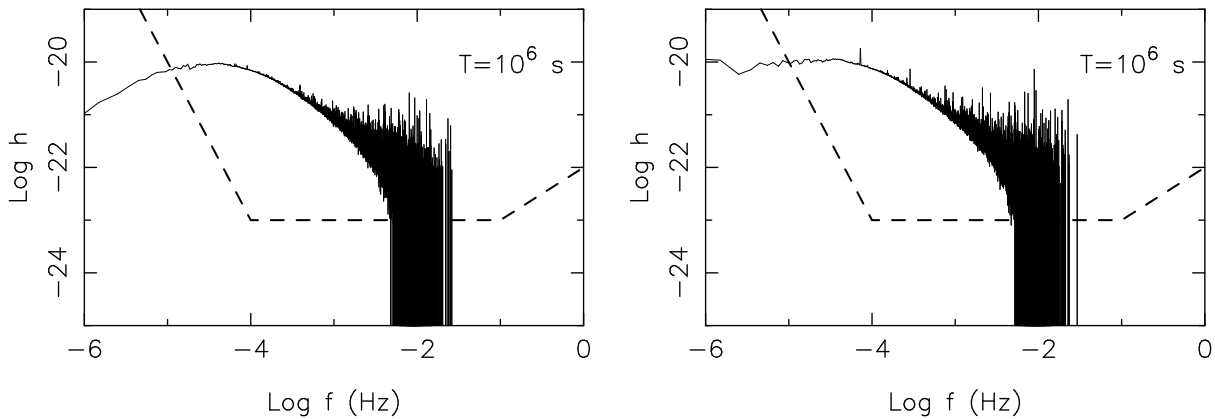

\begin{minipage}{0.4\textwidth}
\psfig{figure=alpha4.ps,width=\textwidth,angle=-90}
\end{minipage}
\hspace*{0.1\textwidth}
\begin{minipage}{0.4\textwidth}
\psfig{figure=all_10e6.ps,width=\textwidth,angle=-90}
\end{minipage}
\caption{Gravitational wave signal for a 10$^6$\,s observation. Left:
power spectrum of double white dwarfs, Right: power spectrum for all
compact binaries in the Galaxy. The planned sensitivity for LISA for
an 10$^6$\,s observation is indicated with the dashed line.}
\label{fig:detection_curves}
\end{figure}

The result for $T = 10^6$\,s is shown in
Figure~\ref{fig:detection_curves} for the population of double white
dwarfs, together with the planned sensitivity for LISA. For
frequencies below $\sim$\,3\,mHz, the number of double white dwarfs is
so large that the signal is confusion limited. This means that in all
frequency bins a large number of double white dwarfs is added
randomly, making it impossible to resolve the individual binaries and
producing a broad noise component.  Only above a few mHz is the number
density of systems sufficiently small that systems can be
resolved individually.  Some double white dwarfs are so close to the
Earth, that their signal is actually strong enough to be detected
above the noise.  The AM CVn systems do not contribute significantly
either to the noise or to the detectable sources.

When we compute the gravitational wave signal for all binaries, there
is no qualitative difference. We still have the double white dwarf
`noise' dominant at low frequencies, but we have some 10 - 100 other
resolved binaries and especially more strong signals above the noise.

\section{Uncertainties and comparison with other authors}

\subsection{Uncertainties}

Our computations are affected by uncertainties in our assumptions
about the parameters of the initial binaries and about the details
of the binary evolution.

The most uncertain of our assumed initial distributions (equation 1)
is that of the mass ratio. In our calculation more than 50\% of all
double white dwarfs is formed from an initial binary with mass ratio
larger than 0.6. An initial distribution that is more skewed towards
higher mass ratios will therefore produce more double white
dwarfs. This difference can be estimated from the work by Yungelson et
al.\,\cite{ylt+94} who use a mass ratio distribution Prob($q)
\propto 2 \; q$ giving a higher merger rate for heavy white dwarfs by
almost a factor 2.

An other uncertainty lies in the normalization of the number of
binaries and single stars in our Galaxy. The star formation history is
inferred from the enrichment of the interstellar gas and involves
detailed modeling of stellar evolution in low metallicity
environments. Note that we also assume a star forming rate which is
constant in time. The value of 4\,$\msun$\,yr$^{-1}$ that we use is
therefore uncertain. Since at least 50\%\ off all stars are in
binaries\,\cite{dm91}, the assumption that all stars are formed in
binaries introduces at most a relatively small error.  We
estimate that the uncertainty in the total number of binaries formed
is about a factor of~2.

The uncertainties in the binary evolution are large especially in the
description the mass transfer and of the spiral-in process in a common
envelope.  Note that at least one common envelope during the evolution
of the binary is necessary for the formation of a double white dwarf
that is close enough to be a significant source of gravitational
waves.

In a common envelope the companion is engulfed by the envelope of the
donor. Frictional forces slow down its motion and heat the envelope
causing it to be expelled. The details of this process involve 3D
hydrodynamics, so we can only guess the efficiency with which the
heating process expels the envelope. Varying this efficiency by a
factor~4 results in birth rates of double white dwarfs which differ by
a factor~2.

When stars transfer matter, but do not get into a common envelope
phase some of the transferred material may be lost, because the
accretor can not accrete all the matter. The angular momentum that is
dragged along with this matter is uncertain.  Varying this angular
momentum by a factor~2 results in birth rates of double white dwarfs
which differ by a factor~5. So this is the most important
uncertainty. In the computations presented here we used a rather high loss
 of angular momentum with the transferred matter (three times the
mean specific angular momentum of the orbit).

\subsection{Comparison with other authors}

\begin{table}[t]
\caption{Birth rates for double white dwarfs by different groups}
\label{tab:birth_rates}
\vspace{0.4cm}
\begin{center}
\begin{tabular}{|l|r|r|}
\hline
  & birth rate Dwd & birth rate AM CVn's \\ 
\hline 
this work                       & 0.1~\,  & 0.0013 \\ 
Han 1998\,\cite{han98}           & 0.03  & 0.026~\, \\ 
Iben et al. 1997\,\cite{ity97}   & 0.09  & 0.02~~~  \\ \hline
\end{tabular}
\end{center}
\end{table}

We compare our results with the work of Han\,\cite{han98} and Iben et
al.\,\cite{ity97} Han uses a grid of
stellar evolution models, computed with the Eggleton
code\,\cite{egg73}. He then models the binary evolution separately and
picks models from the grid. Iben et al. use the binary evolution
program developed by Tutukov and Yungelson and essentially use a
similar approach to ours. They make different assumptions about
the mass transfer phases in the binary evolution, especially when mass
is lost from the system.

As can be seen from Table~\ref{tab:birth_rates} the predicted birth
rates for the double white dwarfs differ by a factor of~5. In part
these differences are due to different assumed normalization of the
initial binary population; in part due to different assumptions about
the loss of angular momentum from the binary.

The discrepancies are even worse for the birth rate of AM CVn systems,
which differ by a factor of~20. The formation of a stable AM CVn
system requires the mass ratio to be less than~2/3.  Both Han and Iben
et al.\ derive a mass ratio distribution of the current double white
dwarf population peaked at~0.5, whereas we find the peak around~0.9
(Figure~\ref{fig:population}). Thus according to Han and Iben et al.\ 
a much larger fraction of double white dwarfs evolves into stable AM
CVn systems. At this moment we do not fully understand why the mass
ratio distributions peak at different values.

\section{Conclusions}

We conclude that the expected population of double white dwarfs
produces a confusion limited noise component in space based
gravitational wave detectors like OMEGA and LISA at frequencies below
$\sim$\,3\,mHz. Above this limit we expect $\gtap 2000$ resolved
binaries, of which the majority are double white dwarfs, but with
a few hundred binaries containing neutron stars and/or black
holes. About 10 - 100 compact binaries (mainly with neutron star or
black hole components) can be detected individually above the noise at
low frequencies.

We further conclude that AM CVn systems are not important sources for
space based detectors, neither as detectable individual sources, nor
as confusion limited noise. 

Note that our computations do not include the formation of binaries in
globular clusters (reviewed in Hut et al.\,\cite{hmg+92}), nor do they
include very exotic binaries which are too rare to be formed in our
limited computed sample of 100,000 binaries.

Finally we stress that uncertainties in population synthesis lead to
different results by different groups. Especially the amount of
angular momentum that is dragged along with matter that leaves the
binary system is uncertain and introduces large differences in the
results. We hope to investigate these differences in the future,
leading to a better understanding of binary evolution and better
predictions of the current population of various types of (compact)
binaries.

\section*{Acknowledgments}
This work was supported by NWO Spinoza grant 08-0 to E.P.J. van den
Heuvel and by NASA through a Hubble Fellowship grant awarded by the
Space Telescope Science Institute, which is operated by AURA Inc., for
NASA under contract NAS\, 5-26555.

\end{document}